\documentclass[]{spie}  

\usepackage{amsmath,amsfonts,amssymb}
\usepackage{graphicx}
\usepackage{mathrsfs}
\usepackage[hidelinks]{hyperref}
\usepackage{lineno}

\title{Advanced Time-Division Multiplexed Readout Chain for the BICEP Array 90/150 GHz Receiver}

\author[a,b]{\href{https://orcid.org/0000-0003-4541-7080}{B.~Cantrall}}%
\author[c]{P.~A.~R.~Ade}%
\author[b,d]{\href{https://orcid.org/0000-0002-9957-448X}{Z.~Ahmed}}%
\author[e]{\href{https://orcid.org/0000-0001-6523-9029}{M.~Amiri}}%
\author[f]{\href{https://orcid.org/0000-0002-8971-1954}{D.~Barkats}}%
\author[g]{\href{https://orcid.org/0000-0002-3351-3078}{R.~Basu~Thakur}}%
\author[h]{\href{https://orcid.org/0000-0001-9185-6514}{C.~A.~Bischoff}}%
\author[a]{\href{https://orcid.org/0000-0003-0848-2756}{D.~Beck}}%
\author[g,i]{J.~J.~Bock}%
\author[j]{V.~Buza}%
\author[g]{\href{https://orcid.org/0000-0002-1630-7854}{J.~R.~Cheshire~IV}}%
\author[k]{J.~Connors}%
\author[l]{\href{https://orcid.org/0000-0002-2088-7345}{J.~Cornelison}}%
\author[m]{M.~Crumrine}%
\author[g]{A.~J.~Cukierman}%
\author[k]{E.~Denison}%
\author[k]{W.B.~Doriese}%
\author[n]{L.~Duband}%
\author[k]{M.~Durkin}%
\author[f]{\href{https://orcid.org/0000-0002-7059-8728}{M.~A.~Echter}}%
\author[o]{\href{https://orcid.org/0009-0007-6718-1730}{M.~Eiben}}%
\author[f,p]{\href{https://orcid.org/0000-0003-4117-6822}{B.~D.~Elwood}}%
\author[g]{\href{https://orcid.org/0000-0002-3790-7314}{S.~Fatigoni}}%
\author[q]{\href{https://orcid.org/0000-0001-8217-6832}{J.~P.~Filippini}}%
\author[a]{A.~Fortes}%
\author[k]{J.~Gard}%
\author[g]{M.~Gao}%
\author[h]{C.~Giannakopoulos}%
\author[a]{N.~Goeckner-Wald}%
\author[a]{\href{https://orcid.org/0000-0001-5268-8423}{D.~C.~Goldfinger}}%
\author[r,s]{S.~Gratton}%
\author[a]{J.~A.~Grayson}%
\author[g]{\href{https://orcid.org/0009-0003-6999-0129}{A.~Greathouse}}%
\author[f]{\href{https://orcid.org/0000-0001-9292-6297}{P.~K.~Grimes}}%
\author[d]{G.~Haller}%
\author[e]{M.~Halpern}%
\author[b,d]{S.~Henderson}%
\author[m]{\href{https://orcid.org/0000-0002-3437-5228}{T.~D.~Hoang}}%
\author[k]{J.~Hubmayr}%
\author[g]{\href{https://orcid.org/0000-0001-5812-1903}{H.~Hui}}%
\author[a]{K.~D.~Irwin}%
\author[t]{M.~Izquierdo~Poza}%
\author[g]{\href{https://orcid.org/0000-0002-3470-2954}{J.~H.~Kang}}%
\author[t]{\href{https://orcid.org/0000-0002-5215-6993}{K.~S.~Karkare}}%
\author[g]{S.~Kefeli}%
\author[f,p]{\href{https://orcid.org/0009-0003-5432-7180}{J.~M.~Kovac}}%
\author[a]{C.~Kuo}%
\author[m,u]{\href{https://orcid.org/0000-0002-4540-1495}{K.~Lasko}}%
\author[g]{\href{https://orcid.org/0000-0002-6445-2407}{K.~Lau}}%
\author[h]{M.~Lautzenhiser}%
\author[a]{\href{https://orcid.org/0000-0001-5677-5188}{T.~Liu}}%
\author[j,v]{\href{https://orcid.org/0000-0002-1414-7236}{S.~C.~Mackey}}%
\author[m]{N.~Maher}%
\author[i]{K.~G.~Megerian}%
\author[g]{L.~Minutolo}%
\author[g]{\href{https://orcid.org/0000-0002-4242-3015}{L.~Moncelsi}}%
\author[a]{Y.~Nakato}%
\author[g,i]{H.~T.~Nguyen}%
\author[g,i]{R.~O’Brient}%
\author[f]{S.~N.~Paine}%
\author[g]{A.~Patel}%
\author[f]{\href{https://orcid.org/0000-0002-4436-4215}{M.~A.~Petroff}}%
\author[f,p]{\href{https://orcid.org/0000-0002-7822-6179}{A.~R.~Polish}}%
\author[n]{T.~Prouve}%
\author[m]{\href{https://orcid.org/0000-0003-3983-6668}{C.~Pryke}}%
\author[k]{C.~D.~Reintsema}%
\author[g]{T.~Romand}%
\author[a]{M.~Salatino}%
\author[d]{L.~Sapozhnikov}%
\author[g]{A.~Schillaci}%
\author[f]{B.~Schmitt}%
\author[b,d]{\href{https://orcid.org/0000-0001-7458-6946}{R.~Shi}}%
\author[m,u]{\href{https://orcid.org/0000-0001-7387-0881}{B.~Singari}}%
\author[g,i]{A.~Soliman}%
\author[f]{T.~St.~Germaine}%
\author[g]{\href{https://orcid.org/0000-0003-0260-605X}{A.~Steiger}}%
\author[g]{B.~Steinbach}%
\author[c]{R.~Sudiwala}%
\author[a,b]{K.~L.~Thompson}%
\author[c]{\href{https://orcid.org/0000-0002-1851-3918}{C.~Tucker}}%
\author[i]{A.~D.~Turner}%
\author[w]{\href{https://orcid.org/0000-0002-3942-1609}{C.~Verg\`{e}s}}%
\author[j,v]{A.~G.~Vieregg}%
\author[g]{\href{https://orcid.org/0000-0002-8232-7343}{A.~Wandui}}%
\author[i]{A.~C.~Weber}%
\author[m]{\href{https://orcid.org/0000-0002-6452-4693}{J.~Willmert}}%
\author[g,b,d]{\href{https://orcid.org/0000-0001-5411-6920}{W.~L.~K.~Wu}}%
\author[a]{H.~Yang}%
\author[j,l]{\href{https://orcid.org/0000-0002-8542-232X}{C.~Yu}}%
\author[f]{\href{https://orcid.org/0000-0001-6924-9072}{L.~Zeng}}%
\author[b]{\href{https://orcid.org/0000-0001-8288-5823}{C.~Zhang}}%
\author[g]{S.~Zhang}%
\affil[a]{Department of Physics, Stanford University, Stanford, CA 94305, USA}%
\affil[b]{Kavli Institute for Particle Astrophysics and Cosmology, Stanford University, Stanford, CA 94305, USA}%
\affil[c]{School of Physics and Astronomy, Cardiff University, Cardiff, CF24 3AA, UK}%
\affil[d]{SLAC National Accelerator Laboratory, Menlo Park, CA 94025, USA}%
\affil[e]{Department of Physics and Astronomy, University of British Columbia, Vancouver, British Columbia, V6T 1Z1, Canada}%
\affil[f]{Center for Astrophysics, Harvard \& Smithsonian, Cambridge, MA 02138, USA}%
\affil[g]{Department of Physics, California Institute of Technology, Pasadena, CA 91125, USA}%
\affil[h]{Department of Physics, University of Cincinnati, Cincinnati, OH 45221, USA}%
\affil[i]{Jet Propulsion Laboratory, California Institute of Technology, Pasadena, CA 91109, USA}%
\affil[j]{Kavli Institute for Cosmological Physics, University of Chicago, Chicago, IL 60637, USA}%
\affil[k]{National Institute of Standards and Technology, Boulder, CO 80305, USA}%
\affil[l]{High-Energy Physics Division, Argonne National Laboratory, Lemont, IL, 60439, USA}%
\affil[m]{School of Physics and Astronomy, University of Minnesota, Minneapolis, MN 55455, USA}%
\affil[n]{Service des Basses Temp\'eratures, Commissariat \`a l'\'Energie Atomique, 38054 Grenoble, France}%
\affil[o]{Faculty of Physical Sciences, University of Iceland, 102 Reykjav\'ik, Iceland}%
\affil[p]{Department of Physics, Harvard University, Cambridge, MA 02138, USA}%
\affil[q]{Department of Physics, University of Illinois at Urbana-Champaign, Urbana, IL 61801, USA}%
\affil[r]{Centre for Theoretical Cosmology, DAMTP, University of Cambridge, Cambridge CB3 0WA, UK}%
\affil[s]{Kavli Institute for Cosmology Cambridge, Cambridge CB3 0HA, UK}%
\affil[t]{Department of Physics, Boston University, Boston, MA 02215, USA}%
\affil[u]{Minnesota Institute for Astrophysics, University of Minnesota, Minneapolis, MN 55455, USA}%
\affil[v]{Department of Physics, University of Chicago, Chicago, IL 60637, USA}%
\affil[w]{Lawrence Berkeley National Laboratory, Berkeley, CA 94720, USA}

\authorinfo{Further author information: (Send correspondence to B.C.)\\B.C.: E-mail: cantrall@stanford.edu\\ }

\begin{document} 
\maketitle

\begin{abstract}
This work presents the current performance of the advanced time-division multiplexed (TDM) readout chain for the BICEP Array 90/150 GHz receiver. BA4-90/150, scheduled for deployment to the South Pole in 2026–27, will use photon-noise-limited, feedhorn-coupled transition edge sensor detectors and an upgraded DC SQUID-based TDM system to map the cosmic microwave background. This new TDM system mitigates readout-induced systematics that are beginning to emerge above the noise floor of the most sensitive maps produced by the BICEP collaboration. Improvements include faster, fully differential SQUID designs, higher TES signal amplification, reduced crosstalk, and hierarchical row-addressing that reduces wiring required for row switching. Measurements made through legacy single-ended warm readout electronics show the upgraded cryogenic readout chain performs as well as or better than the TDM system currently fielded on the BICEP experiment. New warm electronics currently in development at SLAC National Accelerator Laboratory will provide matched fully differential circuits and higher bandwidth, reducing RF susceptibility and aliased noise contributions. On-sky demonstration of this technology will establish a new low-noise, high-bandwidth TDM architecture for future CMB observatories. 
\end{abstract}


\section{INTRODUCTION}
\label{sec:intro}  

The cosmic microwave background (CMB) is the relic radiation released when the Universe cooled sufficiently for radiation and matter to decouple, approximately 380,000 years after the Big Bang. The CMB provides a unique window for studying the physics of the early Universe, and is the primary observable for investigating the theory of cosmic inflation \cite{Linde_1982}. Inflationary theory posits that the Universe underwent a period of extremely rapid expansion in the first $10^{-36}-10^{-32}$ seconds after the Big Bang, increasing in size by roughly $e^{60}$ \cite{Planck_inflation}. This rapid expansion would have generated tensor perturbations, imprinting a stochastic background of primordial gravitational waves. These gravitational waves would induce a parity-odd B-mode polarization signature in the CMB, which, if detected, would provide direct evidence for inflation. 

Using a series of telescopes operated at the South Pole, the BICEP collaboration has set the current world-leading constraint on the tensor-to-scalar ratio $r$, parameterizing the amplitude of the primordial B-mode signal. In combination with Planck and WMAP data, the most recent joint analysis yields $r_{0.05}<0.036$ at 95\% confidence with $\sigma(r)=0.009$ \cite{BK18}. The BICEP3 instrument \cite{BICEP3_xtalk} has been observing for the past decade and has produced the deepest polarization measurements of the CMB at 95 GHz. The BICEP Array (BA) instrument is the current-generation experiment in the BICEP series and, in its final configuration, will feature four BICEP3-style cryogenic receivers observing from 30 to 270 GHz \cite{BA_paper}. Its next installment, BA4-90/150, is a dichroic receiver observing at 90 and 150 GHz that will be deployed to the South Pole experiment site in the 2026--27 austral summer. Both of these instruments employ DC superconducting quantum interference device (SQUID) based time-division multiplexing (TDM) to read out a combined total of $\mathscr{O}(10^4)$ transition edge sensor (TES) detectors across all focal planes. Both the detector and readout technology have been successfully proven by multiple CMB experiments over the past decade \cite{ACTpol, SPIDER, CLASS}. 

Maps from the BICEP collaboration have reached a sensitivity at which readout-induced systematics are beginning to rise above the noise floor, including crosstalk artifacts and bandwidth limitations. These systematics are successfully addressed at the data analysis stage; however, direct hardware mitigations are preferred to avoid complicated analysis procedures and potential residuals in the final data products. To address these effects, the BA4-90/150 receiver will feature the advanced ``mux21'' TDM readout system developed at NIST Boulder \cite{Durkin_twolevel} with higher bandwidth, fully differential signal paths, and hierarchical row-addressing. In addition to upgraded cryogenic readout components, a new suite of TDM warm electronics is currently in development at SLAC National Accelerator Laboratory with deployment on BA4-90/150 planned for the 2027--28 austral summer. This work reports on the current performance of the mux21 cryogenic readout chain for CMB observatories. 

\section{READOUT IMPROVEMENTS IN BA4-90/150}

In traditional TDM readout systems, detectors are arranged in a two-dimensional grid of rows and columns. Each detector is inductively coupled to a first stage SQUID amplifier (SQ1) through an input coil, whose output is in series with the input coil of a common-column second stage SQUID series array amplifier (SSA). The SQ1 is shunted by a flux activated switch, called a row select. When the row select switch is superconducting, it shorts out the SQ1. When an input flux is applied to the row select it is driven normal, activating that row so that each SQ1 reads out its associated detector. Each detector on a given row is read out simultaneously by activating the associated row selects. The sampling rate is thus set by the amount of time it takes to iteratively read out each row in the multiplexing order. 

The mux21 TDM system features three key architectural design changes in comparison to the previous TDM implementation currently in operation on BA, including faster SQUID designs with higher gain and increased bandwidth, improved RF rejection through the implementation of fully differential signal paths \cite{emi}, and hierarchical row addressing that reduces wiring required for row-switching \cite{twolevel-switches}. Implementation of hierarchical row addressing is done through the addition of a second level of flux activated switches, which shunt a subgroup of row selects. Each subgroup of row selects along with its higher level switch is on a dedicated chip of the cryogenic readout. To activate a given row, the corresponding higher level switch must first be activated. The total number of wire pairs required for this switching scheme is the sum of the number of row selects in the subgroup and the number of higher level switches. For example, for 50 rows with a row select subgroup of 10, the required number of wire pairs with mux21 is 15 as opposed to 50 in traditional TDM systems. This reduces the heat load on the lowest temperature stages of the cryogenic system, as well as the integration cost and complexity of these TDM chips. 

The mux21 CMB cryogenic components are currently in the characterization and validation stage, while matched fully differential warm electronics are still under development. To provide a comparison to systems currently operating on BA, measurements in this work are made through the legacy single-ended multi-channel electronics (MCE) \cite{MCE}. All mux21 measurements are made on a readout-only module without a detector wafer, isolating the contributions of the readout chain from TES-mediated thermal and electrical effects.

\subsection{Bandwidth}
\label{sec:title}

\begin{figure}
    \centering
    \includegraphics[width=0.5\linewidth]{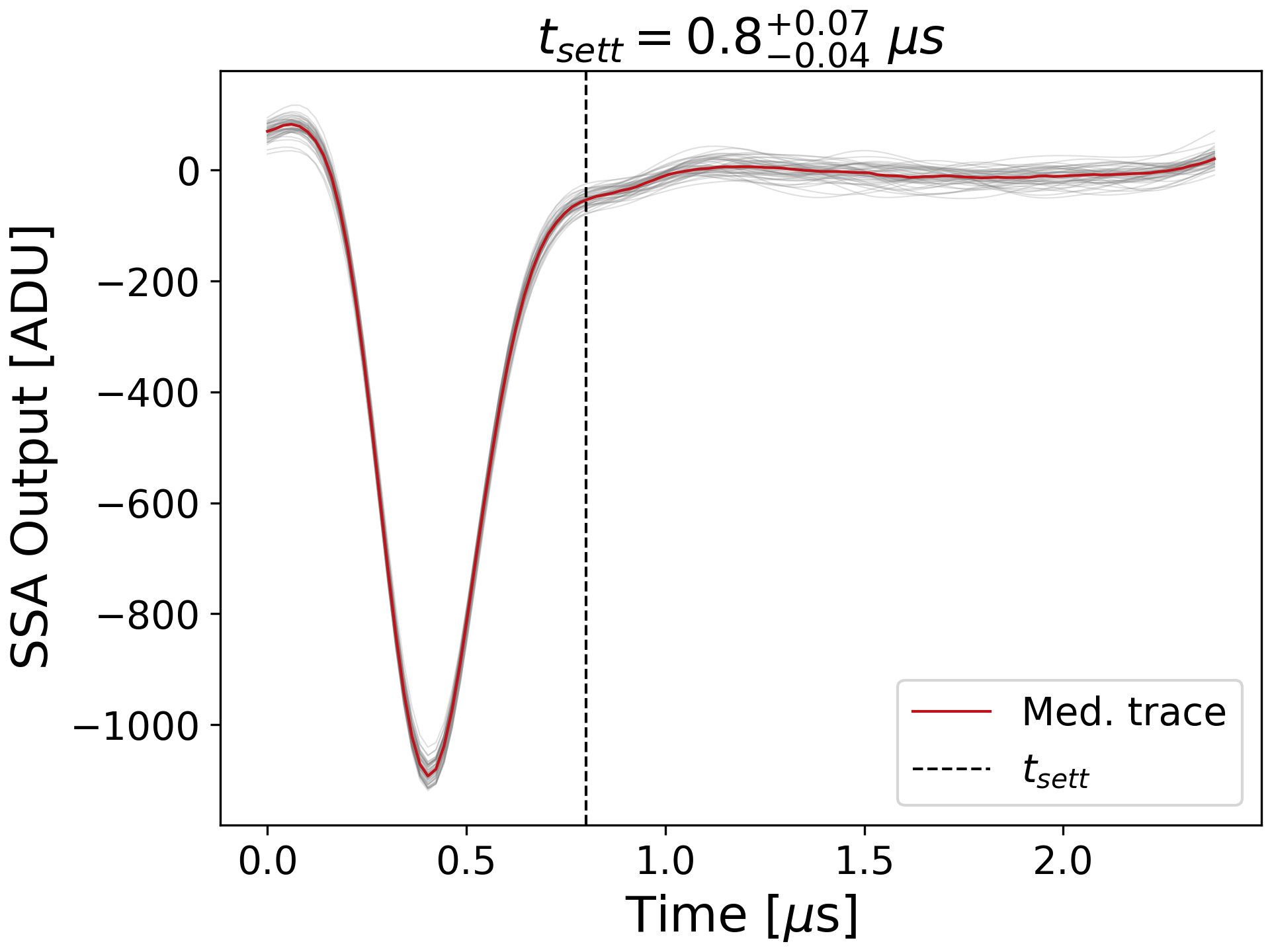}
    \caption{Raw traces of a representative channel of mux21 readout sampled at 50 MHz (gray) and the median across all traces (red). The black dashed vertical line indicates the time at which this channel is determined to be settled after an input flux is applied, $t_\mathrm{sett}$. The title gives the median $t_\mathrm{sett}$ across all traces, with errors given as the $16^\mathrm{th}$ and $84^\mathrm{th}$ percentiles. For this dataset, the row dwelling time was set to 2.4 $\mathrm{\mu s}$ in order to ensure the SQUIDs are fully settled when the row is switched and allow for a robust estimate of settling times.}
    \label{fig:raw_switching_trace}
\end{figure}

High multiplexing factors, or number of rows to be read out in the multiplexing scheme, increase the time required to cycle through all rows for a given row dwelling time. The row dwelling time is set primarily by the cryogenic SQUID components, which have a characteristic response to changes in input flux as shown in Fig. \ref{fig:raw_switching_trace}. The row dwelling time must be long enough for the SQUIDs to settle and allow for readout of the detectors. The sampling frequency of the system is thus set by the multiplexing factor and SQUID settling time and must be high enough to avoid a significant aliased noise penalty, which scales with the square root of the multiplexing factor. 

\begin{figure}
    \centering
    \includegraphics[width=0.75\linewidth]{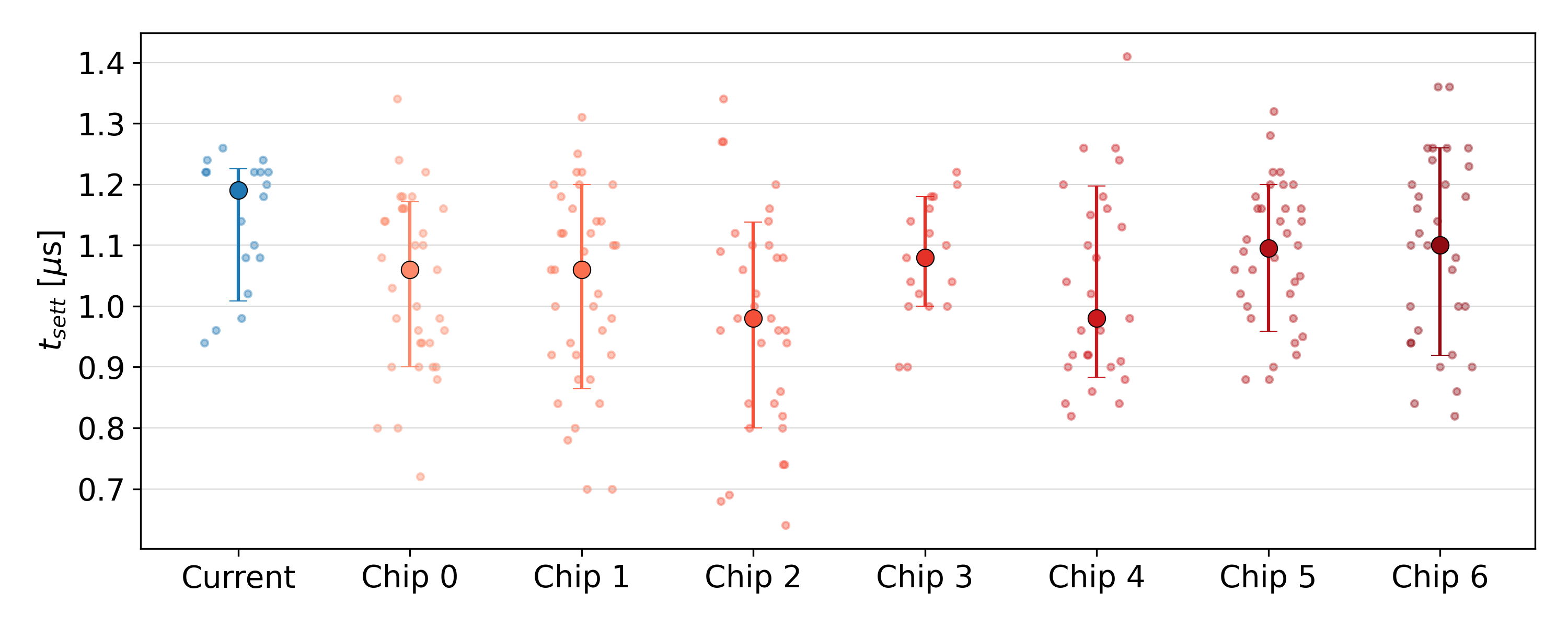}
    \caption{Settling times for 7 row select subgroups (chips) of mux21 readout (red, gradient) and the current TDM implementation (blue). The solid filled points represent the median of the per-component settling times (scattered data points), and error bars indicate the $16^\mathrm{th}$ and $84^\mathrm{th}$ percentiles of the per-component settling times.}
    \label{fig:settle_summary}
\end{figure}

We have measured SQUID settling times on a mux21 readout module through the legacy MCE warm electronics and single-ended SSAs. Results are shown in Fig. \ref{fig:settle_summary} for 7 row select subgroups (chips) of mux21 readout, as well as for the current TDM system. We find a 10\% reduction in median settling times for mux21 cryogenic chips compared to current systems. This is expected to improve when implementing the upgraded warm TDM electronics and faster, fully differential SSAs. 

\subsection{Crosstalk}

\begin{figure}
    \centering
    \includegraphics[width=0.75\linewidth]{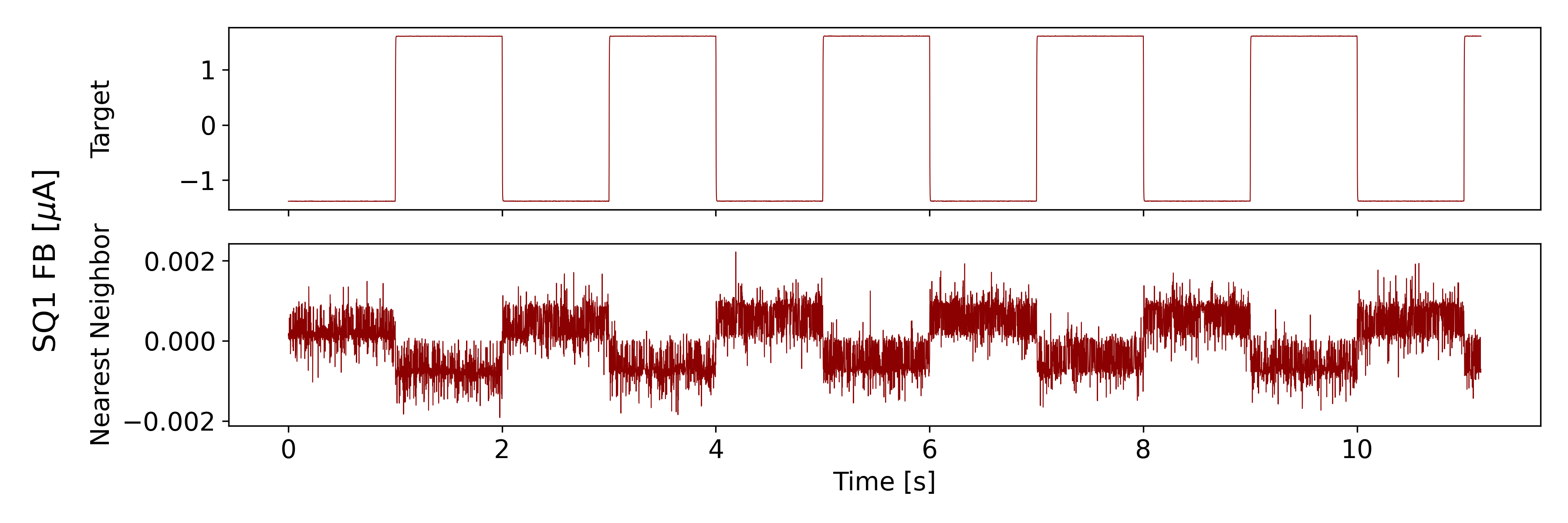}
    \caption{Timestream of a representative channel shorted at the TES input coil of mux21 readout (top) and its time forward nearest neighbor (bottom) sampled at a row dwelling time of 1.78 $\mathrm{\mu s}$. The timestream displays the 0.5 Hz square wave sent down the TES bias line in the Target channel, as well as the perpetrator signal in the Nearest Neighbor channel.}
    \label{fig:xtalk_raw}
\end{figure}

High-density focal planes necessitate close packing of detectors and cryogenic readout components. This close packing can induce channel-to-channel crosstalk in TDM systems through the flux coupling of a SQUID loop with the input coil of a neighboring channel, known as inductive crosstalk. In addition to this inductive component, crosstalk can be introduced through insufficient row settling times. After switching to a new row, a transient current is present in the readout column from the previous row that persists with a time constant set by the self-inductance of the SQ1 input coil and readout circuit resistance. The row dwelling time must be sufficiently longer than this time constant, or detector data will be contaminated with a perpetrator signal from the previous row. This is known as settling time crosstalk. 

We measure time-forward crosstalk in the mux21 readout module by shorting the TES input coil of one row in a given chip. All other rows in the chip are open at the TES input coil. We send a 0.5 Hz square wave down the TES bias line of the readout module. Only channels that are shorted at the TES input coil should show a readout response to this square wave. The row shorted at the TES input coil is iterated through the multiplexing order so that a measurement of settling time crosstalk can be made for every open row. An example timestream for a representative shorted channel and its time nearest-neighbor is shown in Fig. \ref{fig:xtalk_raw}.

\begin{figure}
    \centering
    \includegraphics[width=0.5\linewidth]{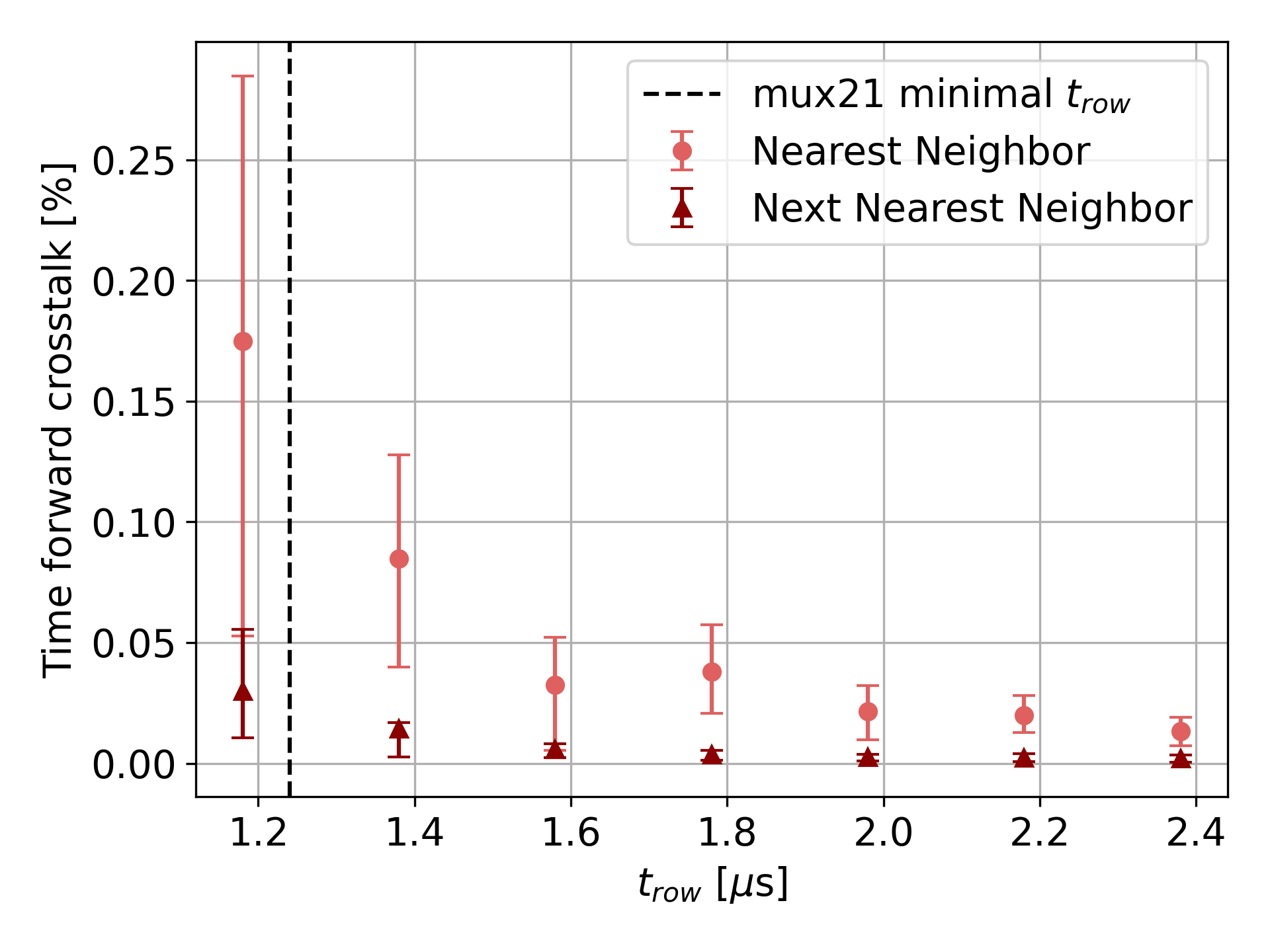}
    \caption{Percent amplitude crosstalk in the time forward nearest- (circle) and next-nearest-neighbors (triangle) of channels shorted at the TES input coil as a function of row dwelling time $t_\mathrm{row}$. Error bars indicate the $16^\mathrm{th}$ and $84^\mathrm{th}$ percentiles of the datasets at each $t_\mathrm{row}$. The minimal $t_\mathrm{row}$ shown by the vertical black dashed line is determined by the median settling time from Fig. \ref{fig:settle_summary}, with 0.2 $\mathrm{\mu s}$ added to account for the acquisition of detector data.}
    \label{fig:xtalk_multistep}
\end{figure}

The percent amplitude of the square wave present in the time nearest-neighbor and next-nearest-neighbor channel is measured as a function of row dwelling time $t_\mathrm{row}$. Results are shown in Fig. \ref{fig:xtalk_multistep}. For $t_\mathrm{row}>2$ $\mathrm{\mu s}$, we measure a mean crosstalk amplitude of $\sim$0.02\%. We interpolate the crosstalk percent amplitude at the minimal $t_\mathrm{row}$ of 1.24 $\mathrm{\mu s}$, taken as the median settling time from Fig. \ref{fig:settle_summary} with 0.2 $\mathrm{\mu s}$ added to account for the acquisition of detector data, at 0.14\%. Previous measurements of crosstalk on the BICEP3 instrument yielded 0.3\% \cite{BICEP3_xtalk}. We measure an order of magnitude reduction in the percent amplitude crosstalk for equivalent BICEP3 $t_\mathrm{row}$.  

\subsection{Differential Signaling}

\begin{figure}
    \centering
    \includegraphics[width=0.5\linewidth]{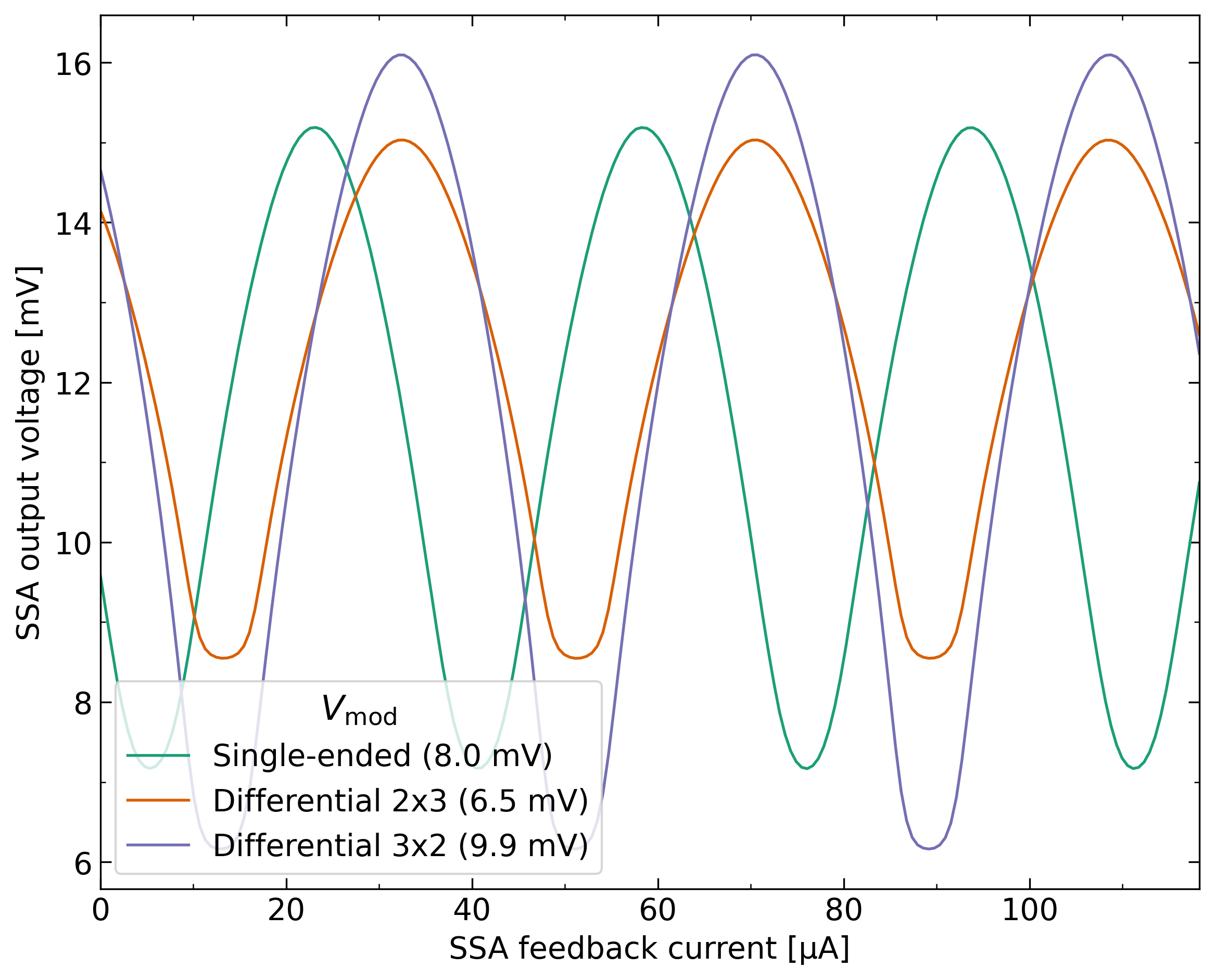}
    \caption{Modulation curves of the single-ended SSA devices (green) and two choices of differential SSAs. The 3x2 SSAs (purple) showcase increased gain at the cost of a reduction in bandwidth, while the 2x3 SSAs (red) have increased bandwidth with reduced gain. The gain is directly related to the modulation depth of the curves, shown in the legend as $V_\mathrm{mod}$.}
    \label{fig:ssa}
\end{figure}

We present modulation curves of the first fully differential SSA modules for CMB readout in Fig. \ref{fig:ssa}. There are two implementations of the differential SSA modules: 2-series by 3-parallel (2x3) and 3-series by 2-parallel (3x2). The 2x3 SSAs are designed with reduced gain to increase the bandwidth of the system, while the 3x2 SSAs have increased gain with reduced bandwidth. This difference is demonstrated by $V_\mathrm{mod}$, the peak-to-peak amplitude of the modulation curve, for each SSA shown in Fig. \ref{fig:ssa}, where the 3x2 has the greatest modulation depth of 9.9 mV. This increase in SSA gain reduces the readout noise contribution referred to the TES input. The selection of SSA to be deployed in the BA4-90/150 GHz receiver is currently under consideration. Demonstration of the improvement in RF rejection of the differential SSAs must wait until the upgraded warm electronics are ready. 

\section{Conclusions}
\label{sec:sections}

This work presents the current performance of the upgraded TDM system for CMB experiments, to be deployed for the first time at the South Pole on the BA4-90/150 GHz receiver in the 2026--27 austral summer. This system implements architectural changes that address readout systematics currently affecting the most sensitive maps produced by the BICEP experiment. We measure a 10\% reduction in SQUID settling times, demonstrating the increase in bandwidth through faster SQUID designs that will improve when moving to the new TDM warm electronics currently in development at SLAC National Accelerator Laboratory. We find a crosstalk percent amplitude an order of magnitude smaller than previous measurements made on the BICEP3 instrument at equivalent row dwelling times. Finally, we demonstrate the first operation of fully differential SSAs that will enable improvements in the RF rejection of the readout chain. The upgraded warm electronics are planned for deployment on this receiver in the 2027--28 austral summer, completing the advanced TDM readout chain. Demonstration of this system on BA4-90/150 will establish a new low-noise, high-bandwidth TDM architecture for future CMB experiments.


\bibliography{report} 
\bibliographystyle{spiebib} 

\end{document}